\documentclass[conference]{IEEEtran}

\usepackage[cmex10]{amsmath}
\usepackage{amssymb}

\usepackage{cases}
\usepackage{multirow}
\usepackage{empheq}

\usepackage{units}
\usepackage{cite}
\usepackage{verbatim}

\usepackage{algorithm}
\usepackage{algpseudocode}

\usepackage{siunitx}

\ifCLASSINFOpdf
\usepackage[pdftex]{graphicx}
\else
\usepackage[dvips]{graphicx}
\fi
\usepackage[caption=false,font=footnotesize]{subfig}

\usepackage{color}
\usepackage{multirow}

\newtheorem{theorem}{Theorem}

\newtheorem{remark}[theorem]{Remark}

\newcommand{\vect}[1]{\mathbf{#1}}

\begin{document}
	
	\sloppy
	
	\title{On the Optimal Refresh Power Allocation for Energy-Efficient Memories}
	
	
	\author{\IEEEauthorblockN{Yongjune Kim\IEEEauthorrefmark{1},
			Won Ho Choi\IEEEauthorrefmark{1},		
			Cyril Guyot\IEEEauthorrefmark{1}, and 
			Yuval Cassuto\IEEEauthorrefmark{1}\IEEEauthorrefmark{2}}
		
		\IEEEauthorblockA{\IEEEauthorrefmark{1}Western Digital Research, Milpitas, CA, USA \\ Email: \{yongjune.kim, won.ho.choi, cyril.guyot\}@wdc.com}
		
		\IEEEauthorblockA{\IEEEauthorrefmark{2}Viterbi Department of Electrical Engineering, Technion -- Israel Institute of Technology, Haifa, Israel \\
			Email: ycassuto@ee.technion.ac.il}
	}

	
	\maketitle
	
	\begin{abstract}
		Refresh is an important operation to prevent loss of data in dynamic random-access memory (DRAM). However, frequent refresh operations incur considerable power consumption and degrade system performance. Refresh power cost is especially significant in high-capacity memory devices and battery-powered edge/mobile applications.
		In this paper, we propose a principled approach to optimizing the refresh power allocation. Given a model for the bit error rate dependence on power, we formulate a convex optimization problem to minimize the word mean squared error for a refresh power constraint; hence we can guarantee the optimality of the obtained refresh power allocations. In addition, we provide an integer programming problem to optimize the discrete refresh interval assignments. For an 8-bit accessed word, numerical results show that the optimized nonuniform refresh intervals reduce the refresh power by \SI{29}{\%} at a peak signal-to-noise ratio of \SI{50}{dB} compared to the uniform assignment. 
	\end{abstract}
	
	\section{Introduction}
	
	Memory refresh is a periodically repeated procedure that reads and rewrites the data of a memory device to prevent loss of data. It is well known that dynamic random-access memory (DRAM) cells must be refreshed periodically due to charge leakage~\cite{Bhati2016dram,Emma2008rethinking}. A DRAM cell stores one bit of information by controlling the amount of charge on its capacitor. DRAM cells cannot retain their data permanently because of the gradual loss of charge over time. The time a cell can retain its data is called the \emph{retention time} of the cell. The time interval between refresh operations is the \emph{refresh interval}, which is the inverse of the \emph{refresh rate}. A cell that cannot retain its data for the given refresh interval results in a failure, referred to as \emph{retention failure (or retention error)}~\cite{Ohsawa1998optimizing,Liu2012raidr,Khan2014efficacy}. The typical refresh interval in current DRAM standards is \SI{64}{\milli\second}, which is a conservative value~\cite{Liu2012raidr,Khan2014efficacy}.  
	
	The conservative refresh operations lead to high refresh power consumption. This problem is expected to worsen as DRAM device capacity increases~\cite{Bhati2016dram,Liu2012raidr}. As cell dimension shrinks, memory cells become susceptible to charge leakage and require more frequent refresh operations~\cite{Khan2014efficacy}. Further, the refresh power consumption is critical in battery-powered edge/mobile computing applications. Note that edge/mobile devices are idle most of the time and refresh operations are still required during idle periods unlike write and read operations~\cite{Liu2011flikker}.      
	
	Many refresh techniques were proposed to reduce refresh power~\cite{Ohsawa1998optimizing,Ghosh2007smart,Katayama1999fault,Emma2008rethinking,Liu2012raidr,Khan2014efficacy,Liu2011flikker,Wilkerson2010reducing,Chou2015reducing,Cho2014edram}. Ohsawa et al.~\cite{Ohsawa1998optimizing} and Ghosh et al.~\cite{Ghosh2007smart} proposed architectural techniques to avoid unnecessary refresh operations. Error control coding (ECC) schemes were proposed to decrease refresh rates and correct the resulting retention failures~\cite{Katayama1999fault,Chou2015reducing,Wilkerson2010reducing,Emma2008rethinking}. These ECC schemes suffer from storage or bandwidth overheads. RAIDR~\cite{Liu2012raidr} allocates different refresh intervals by identifying weak DRAM cells. Flikker~\cite{Liu2011flikker} specifies critical and non-critical data and refreshes the memory cells storing non-critical data at a lower rate. Cho et al.~\cite{Cho2014edram} proposed tiered-reliability memory (TRM) to allocate different refresh intervals depending on the importance of bit positions. Since these previous techniques choose the refresh intervals empirically, the granularity of refresh interval assignments are inherently limited. Further, the optimality of refresh intervals has not been addressed. 
	
	We note that refresh is also considered in storage-class memories such as magnetic RAMs (MRAMs) and resistive RAMs (ReRAMs)~\cite{Choi2018comprehensive}. For example, MRAMs suffer from high write latency and energy, which are the key drawbacks of MRAM technology. Several techniques~\cite{Smullen2011relaxing,Jog2012cache} attempt to address the write-inefficiency of MRAMs via relaxing retention time and introducing refresh operations. For the sake of concreteness, we focus on DRAM refresh, wherein refresh has been established as a central trade-off between power and fidelity.  
	
	This paper presents a \emph{principled} approach to refresh interval assignments for machine learning (ML) and signal processing tasks. In these applications, the mean squared error (MSE) is a more meaningful fidelity metric than the bit error rate (BER). We formulate a convex optimization problem to minimize the MSE for a given refresh power constraint. Since the formulated problem is convex, the global optimal solutions can be obtained with standard convex programming algorithms. Even more favorably, we derive an analytic expression for the optimal solution using the Karush-Kuhn-Tucker (KKT) conditions. In addition, we formulate a discrete optimization problem by taking into account hardware implementation. Our evaluation shows that the penalty due to discrete intervals is marginal. A prior study in~\cite{Kim2018generalized} of voltage-swing optimization in static RAMs (SRAMs) is similar in spirit, but its results are not applicable to optimizing DRAM's refresh intervals. To the best of our knowledge, our work is the first rigorous treatment of the optimal refresh
	interval assignments, viz. refresh power allocations.
	
	The rest of this paper is organized as follows. Section~\ref{sec:dram} explains the current DRAM architecture and refresh operations. Section~\ref{sec:model} introduces the optimization metrics of DRAM's refresh power and fidelity. Section~\ref{sec:optimization} formulates optimization problems to determine the optimum refresh intervals and provides the theoretical analysis. Section~\ref{sec:numerical} gives numerical results and Section~\ref{sec:conclusion} concludes. 
	
	\section{DRAM Architecture and Refresh Operations}\label{sec:dram}
	
	\subsection{DRAM Architecture}
	
\begin{figure}[t]
		\centering
		\subfloat[]{\includegraphics[width=0.28\textwidth]{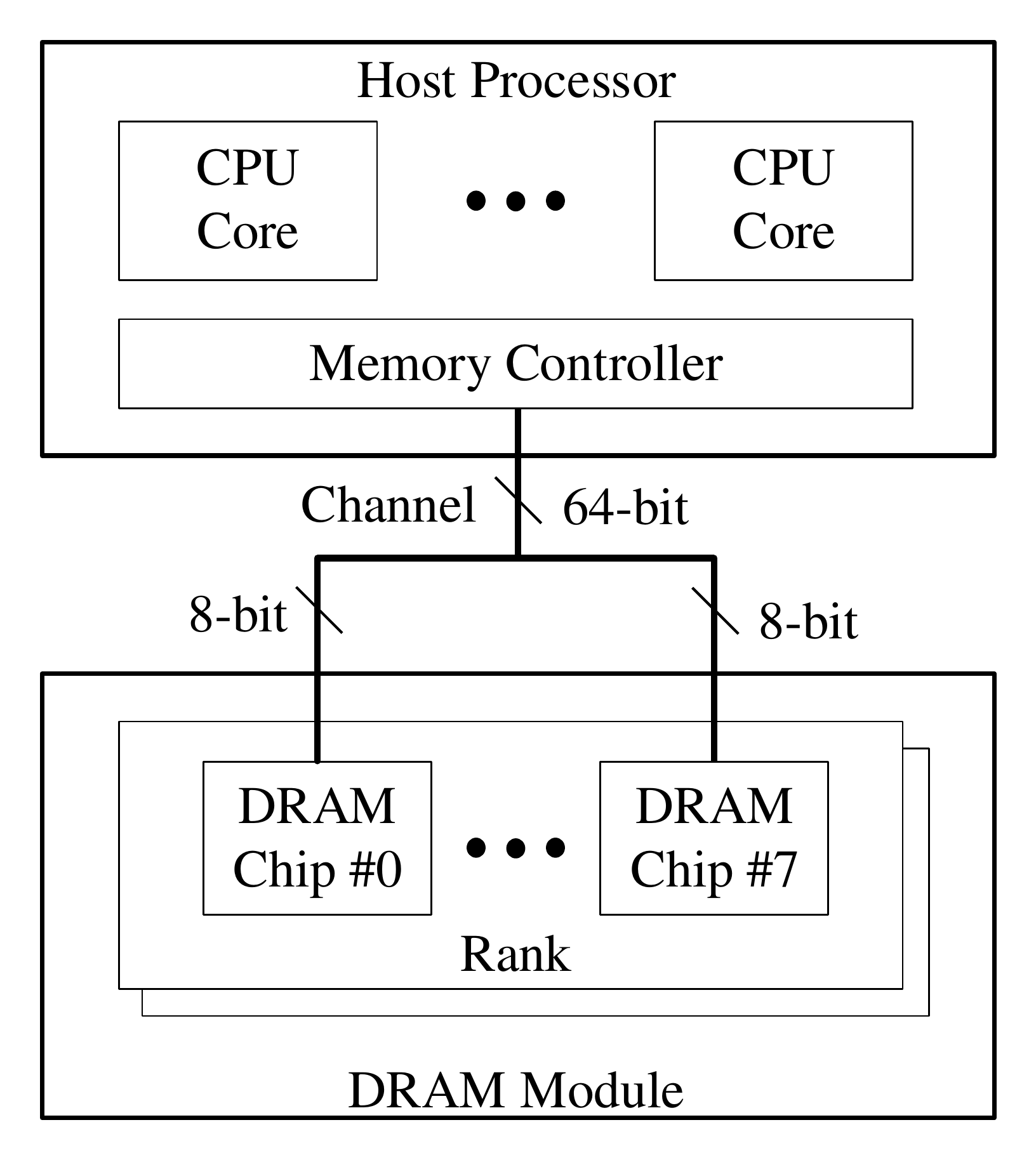}
			\label{fig:dram_arch}}
		\hfil
		\subfloat[]{\includegraphics[width=0.28\textwidth]{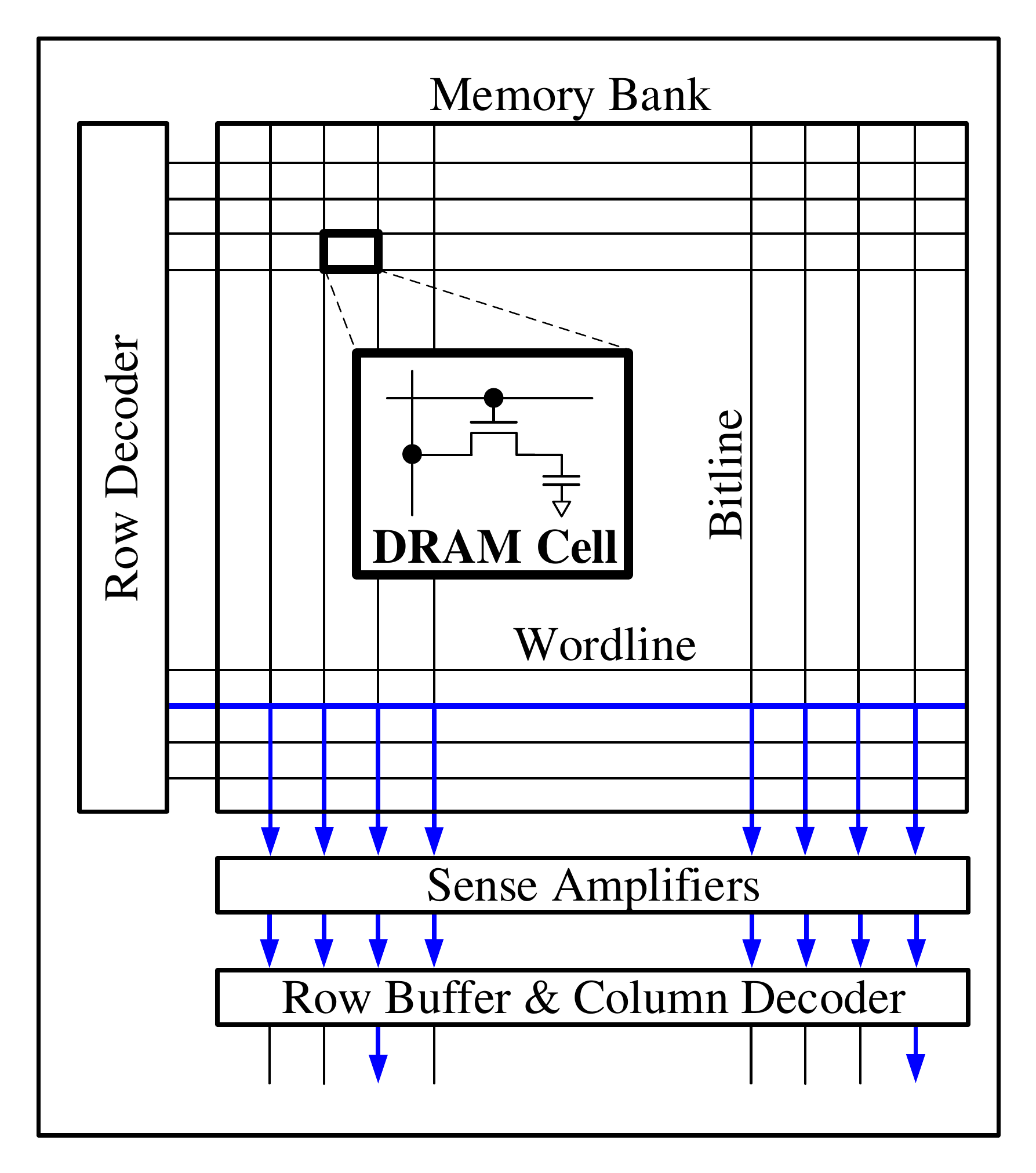}
			\vspace{-4mm}
			\label{fig:dram_chip}}
		\caption{Organization of DRAM system: (a) DRAM system and (b) DRAM bank architecture.}
		\label{fig:dram}
		\vspace{-4mm}
	\end{figure}

	DRAM system is hierarchically organized channels, modules, ranks, and chips as shown in Fig.~\ref{fig:dram}. Each memory channel drives commands, addresses, and data between a memory controller and one or more DRAM modules~\cite{Khan2014efficacy,Chang2016understanding}. Each module contains multiple DRAM chips that are organized into one or more ranks. A rank consists of multiple chips that operate synchronously to provide a wide data bus (e.g., 64-bit) to increase the bandwidth, as a single DRAM chip is designed to have a narrow data bus width (e.g., 8-bit)~\cite{Chang2016understanding}. Each of the eight chips in the rank transfers 8 bits simultaneously in a unit interval of double-data rate (DDR) time frame to provide 64 bits of data as shown in Fig.~\ref{fig:dram}\subref{fig:dram_arch}.
	
	A DRAM chip consists of multiple banks that can process DRAM commands independently to increase parallelism. A bank includes a memory array of DRAM cells that are organized into rows and columns, as shown in Fig.~\ref{fig:dram}\subref{fig:dram_chip}~\cite{Chang2016understanding}. A row consists of \SI{1}{KB} or \SI{2}{KB} cells in general and the number of rows depends on the chip capacity.
	
	A cell has (i) a capacitor that stores binary data in the form of stored charge (e.g., charged and discharged states compared to a reference charge represent 1 and 0, respectively), and (ii) an access transistor that serves as a voltage-controlled switch to connect the capacitor to the bitline~\cite{Khan2014efficacy,Chang2016understanding}. DRAM cells in each column share a bitline, which connects them to a sense amplifier. The sense amplifier detects the charge stored in a cell and converts the charge to binary information. DRAM cells in each row share a wire called the wordline, which controls the corresponding cells' access transistors. When a wordline is enabled by the row decoder, the entire cells in the row get connected to the sense amplifiers through the bitlines, enabling the sense amplifiers to detect the data and latch them into the row buffer~\cite{Chang2016understanding}. A chunk of the data in the row buffer is fetched out by the column decoder. 
	
	\subsection{Refresh Operations}
	
	Since a DRAM cell capacitor leaks charge over time, the charge on each capacitor must be periodically refreshed. To prevent retention failure, the refresh interval should be less than the retention time. Since all memory cells do not have the same retention time because of process variations~\cite{Bhati2016dram,Liu2012raidr,Liu2013experimental}, the BER due to retention failure is given by
	\begin{equation}\label{eq:retention}
	p = \Pr\left(T_{\text{retention}} < t\right),
	\end{equation}
	where $t$ denotes a given refresh interval value. The random variable  $T_{\text{retention}}$ represents the retention time of DRAM cells. It is clear that shorter refresh intervals decrease the BER due to retention failure. To guarantee data integrity, current DRAM standards conservatively employ the refresh interval of \SI{64}{\milli\second}. 
	
	The refresh power $P$ is inversely proportional to the refresh interval as follows~\cite{Liu2011flikker,Kim2003block}:
	\begin{equation}\label{eq:power}
	P \propto \frac{C}{t},
	\end{equation}
	where $C$ denotes the effective switching capacitance. This effective switching capacitance increases for higher-capacity DRAM devices. Hence, the refresh power consumption continues to increase as DRAM device capacity increases~\cite{Liu2012raidr,Bhati2016dram,Kim2003block}.       
	
	
	\section{DRAM Optimization Metrics}\label{sec:model}
	
\begin{figure}[t]
		\centering
		\includegraphics[width=0.45\textwidth]{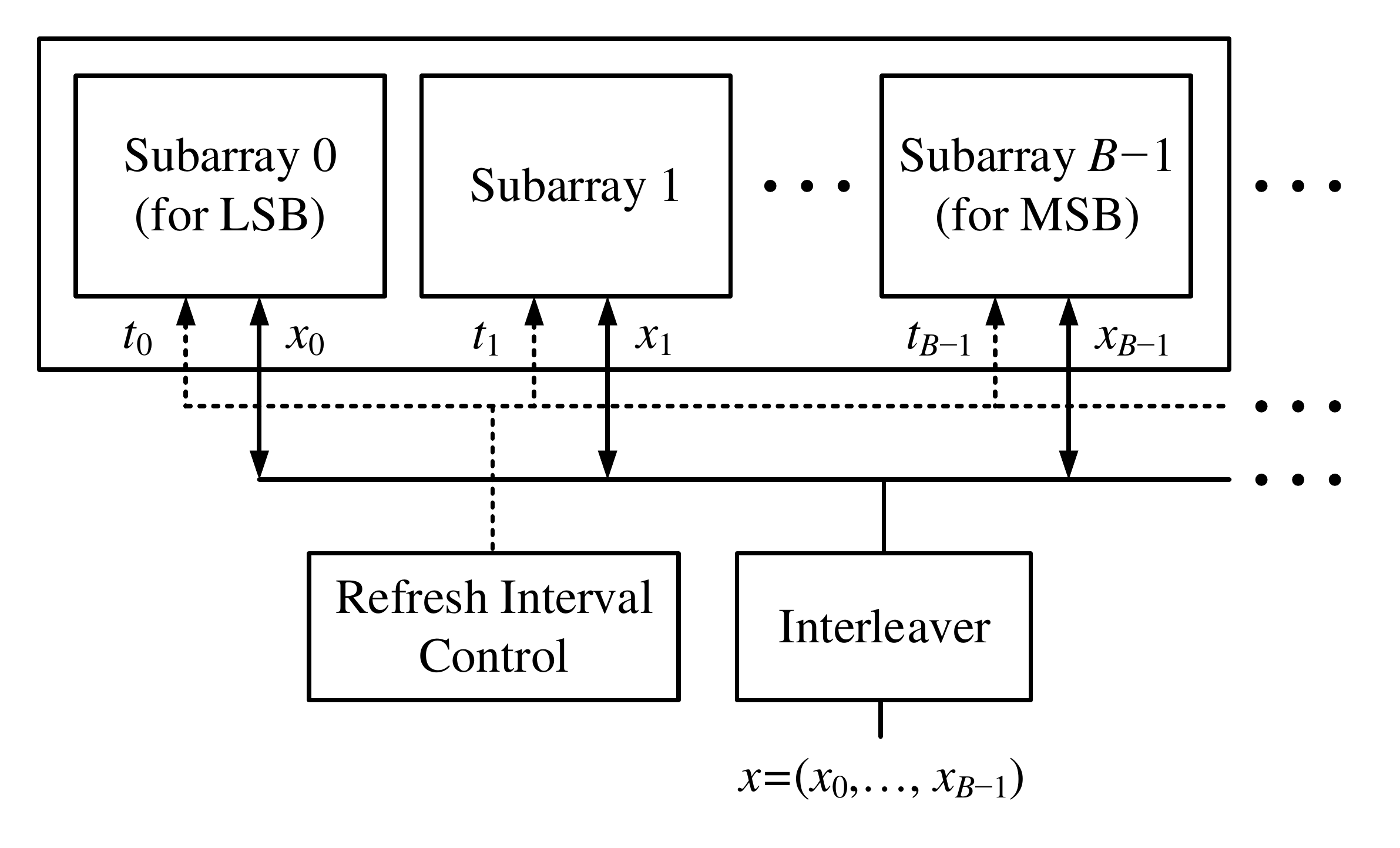}
		\vspace{-4mm}
		\caption{Interleaved architecture~\cite{Cho2014edram} where $x = (x_0, \ldots, x_{B-1})$ denotes a stored $B$-bit word.}
		\label{fig:arch}
		\vspace{-4mm}
	\end{figure}

	The refresh interval $t$ is a key parameter to control the trade-off between refresh power and fidelity. If we separate the data for each bit position in different subarrays by interleaving as in~\cite{Cho2014edram,Kim2018generalized,Kim2018sram}, then the corresponding refresh interval assignment is represented by a vector $\vect{t} = (t_0, \ldots, t_{B-1})$ as shown in Fig.~\ref{fig:arch}. Note that $t_0$ and $t_{B-1}$ represent the refresh intervals corresponding to least significant bit (LSB) and most significant bit (MSB), respectively. Subarrays can correspond to memory banks or memory chips depending on architecture configuration. Due to the current DRAM's multi-chip and multi-bank architecture in Fig.~\ref{fig:dram}, we can allocate different refresh intervals to each subarray with minimal hardware overhead~\cite{Liu2012raidr,Liu2011flikker,Cho2014edram}.  
	
	In the following subsections, we describe the resource and fidelity metrics with the refresh interval assignment. 
	
	\subsection{Resource Metric: Refresh Power}
	
	From~\eqref{eq:power}, the normalized refresh power for a $B$-bit word is given by
	\begin{equation}\label{eq:refresh_power}
	\mathsf{P}(\vect{t}) = \sum_{b=0}^{B-1}{\frac{1}{t_b}}. 
	\end{equation}  
	
	\begin{remark}\label{remark:refresh_power}
		The refresh power $\mathsf{P}(\vect{t})$ is a \emph{convex} function of $\vect{t}$ because $t_b > 0$ for $b \in [0, B-1]$.  
	\end{remark}
	
	\subsection{Fidelity Metrics: BER and MSE}
	
	Suppose that $p_b$ denotes the BER of the $b$th bit position. Since $p_b$ is a function of refresh interval $t_b$, we set
	\begin{equation}
	p_b = g(t_b)
	\end{equation}
	for $b \in [0, B-1]$.  
	
	In many signal processing and ML tasks, the impact of bit errors depends on the bit position. For example, errors in the MSB position of image pixels degrade overall image quality much more than errors in the LSB position. Likely, an MSB error can cause a catastrophic loss in the inference accuracy of ML applications~\cite{Kim2018generalized}. Hence, we use the MSE as a fidelity metric instead of the BER. 
	
	The MSE of $B$-bit words is given by
	\begin{equation}\label{eq:mse}
	\mathsf{MSE}(\vect{t}) = \sum_{b=0}^{B-1}{4^b g(t_b)}, 
	\end{equation} 
	where the weight $4^b$ represents the differential importance of each bit position~\cite{Yang2011unequal,Kim2018generalized}. 
	
	\begin{remark}
		$\mathsf{MSE}(\vect{t})$ is \emph{convex} if $g(\cdot)$ is convex. It is because a nonnegative weighted sum of convex functions is convex.
	\end{remark}
	
	It was reported that the BER increases exponentially with the refresh interval~\cite{Khan2014efficacy,Liu2011flikker,Hou2013fpga,Cho2014edram}. Hence, we model the BER as 
	\begin{equation}\label{eq:exponential}
	p_b = g(t_b) =  \alpha \exp(\beta t_b),
	\end{equation}
	where positive values of $\alpha$ and $\beta$ depend on the memory fabrication parameters.  
	
	\begin{remark}\label{remark:mse_exponential}
		$\mathsf{MSE}(\vect{t})$ is \emph{convex} if $g(\cdot)$ is an exponential function as in \eqref{eq:exponential}. 
	\end{remark}

	\begin{table}[!t]
		\renewcommand{\arraystretch}{1.4}
		\caption{Resource and Fidelity Metrics for Refresh Operation}
		\vspace{-2mm}
		\label{tab:comparison}
		\centering
		\begin{tabular}{|c|c|c|}	\hline
			&  Single bit  & $B$-bit word \\ \hline \hline
			Variable & $t$  &  $\vect{t} = (t_0, \ldots, t_{B-1})$ \\ \hline
			Refresh power   & $\frac{1}{t}$  & $\sum_{b=0}^{B-1}{\frac{1}{t_b}}$ \\ \hline
			Fidelity & $g(t)$ & $ \sum_{b=0}^{B-1}{4^b g(t_b)}$ \\ \hline			
		\end{tabular}
		\vspace{-4mm}
	\end{table} 
	
	Table~\ref{tab:comparison} summarizes the resource and fidelity metrics for single-bit and $B$-bit word. We note that these metrics are convex.  
	
	\section{Formulation of Optimization Problems}\label{sec:optimization}
	
	\subsection{Convex Optimization Problem}
	
	We formulate a convex optimization problem to determine the optimal refresh intervals. For a given refresh power constraint, we seek to minimize MSE as follows: 
	
	\begin{equation}
	\begin{aligned} \label{eq:min_mse}
	& \underset{\vect{t}}{\text{minimize}}
	& & \mathsf{MSE}(\vect{t}) = \sum_{b=0}^{B-1}{4^b \alpha \exp(\beta t_b)}  \\
	&{\text{subject~to}} & & \mathsf{P}(\vect{t}) = \sum_{b=0}^{B-1}{\frac{1}{t_b}} \le \mathcal{P} \\
	& & & t_b \ge \delta, \quad b=0,\ldots,B-1
	\end{aligned}
	\end{equation}
	where $\mathcal{P}$ is a constant corresponding to the given refresh power budget. Note that $\delta > 0$ denotes the conservative minimum refresh interval, which in particular prevents $t_b = 0$ (i.e., infinite refresh power). We set $\delta = 0.064$ based on current DRAM standards.   
	
	Because of Remark~\ref{remark:refresh_power} and Remark~\ref{remark:mse_exponential}, the optimization problem \eqref{eq:min_mse} is convex. Hence, we can obtain the global optimal solutions by standard convex programming algorithms. In addition, we can derive the optimal solution based on KKT conditions.      
	
	\begin{theorem}\label{thm:min_mse}
		The optimal refresh-interval vector $\vect{t}^*$ of \eqref{eq:min_mse} is given by 
		\begin{equation}\label{eq:min_mse_sol}
		t_b^* =
		\begin{cases}
		\delta, & \text{if}\: \frac{\nu}{4^b} < \alpha \beta \delta^2 \exp(\beta \delta); \\
		\frac{2}{\beta} W\left( \frac{\beta}{2} \sqrt{\frac{\nu}{4^b \alpha \beta}} \right), & \text{otherwise}
		\end{cases}
		\end{equation}
		where $\nu$ is a dual variable of KKT conditions. Note that $\nu$ depends on the refresh power budget $\mathcal{P}$ for the given $\alpha$ and $\beta$. We can find $\nu$ efficiently by the bisection method as in~\cite{Palomar2005practical}. Also, $W(\cdot)$ denotes the \emph{Lambert W function}, which is the inverse function of $f(x) = x e^x$~\cite{Corless1996lambert}. 
		
	\end{theorem}
	\begin{IEEEproof}
		We define the Lagrangian $L_1(\vect{t}, \nu, \vect{\lambda})$ associated with problem \eqref{eq:min_mse} as
		\begin{align}
		L_1(\vect{t}, \nu, \vect{\lambda}) & = \sum_{b=0}^{B-1}{4^b \alpha \exp(\beta t_b)} \nonumber \\
		&+ \nu \left( \sum_{b=0}^{B-1}{\frac{1}{t_b} - \mathcal{P}}\right) - \sum_{b=0}^{B-1}{\lambda_b \left(t_b - \delta \right)} 	
		\end{align}
		where $\nu$ and $\vect{\lambda}=(\lambda_0,\ldots,\lambda_{B-1})$ are the dual variables. The optimal solution is derived from $L_1$ and the corresponding KKT conditions. The details of the proof are given in Appendix~\ref{pf:min_mse}.
	\end{IEEEproof}
	
	\begin{figure}[t]
		\centering
		\includegraphics[width=0.36\textwidth]{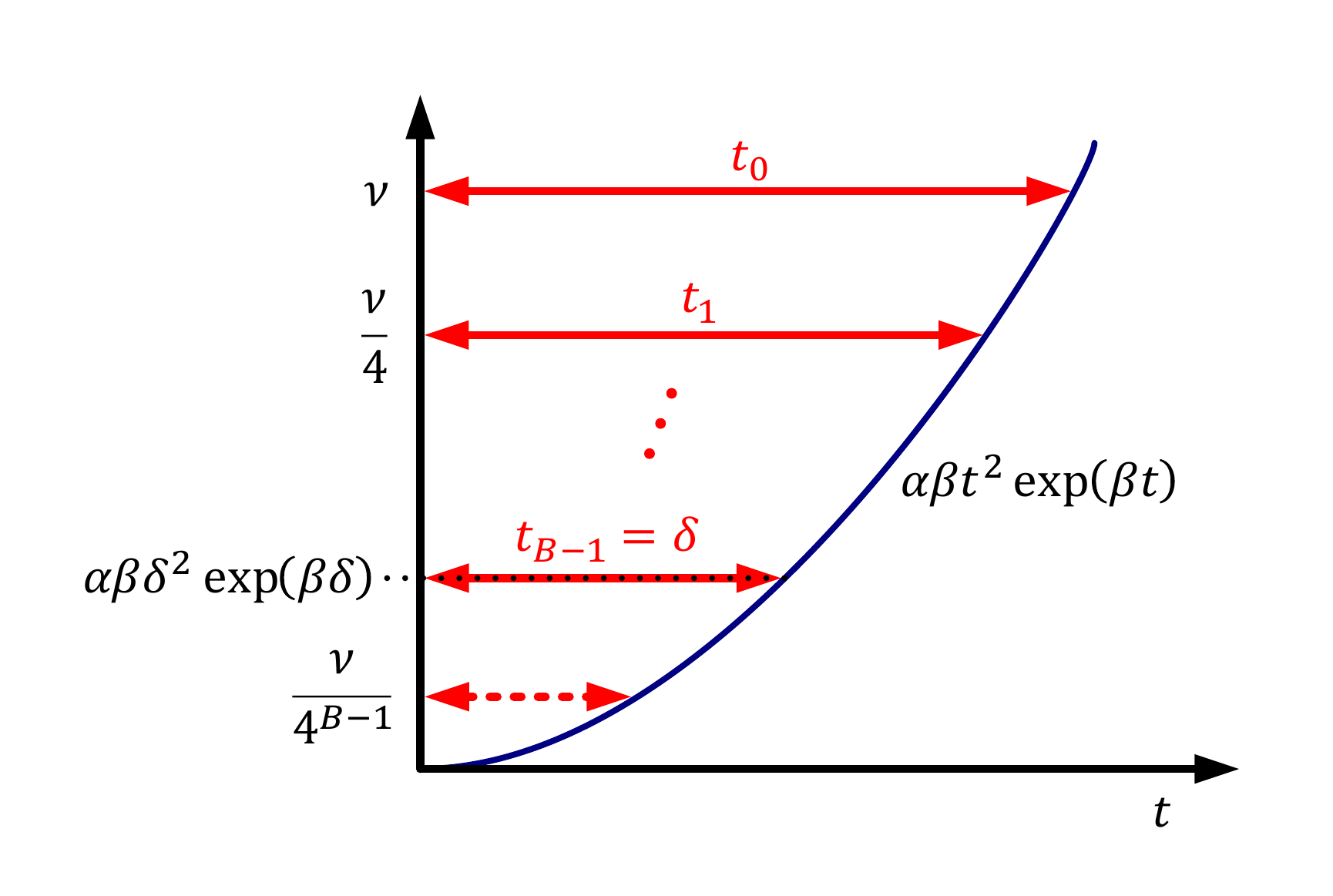}
		\vspace{-3mm}
		\caption{A graphical interpretation of the optimal refresh intervals in Theorem~\ref{thm:min_mse}.}
		\label{fig:solution}
		\vspace{-4mm}
	\end{figure}

	The optimal refresh interval \eqref{eq:min_mse_sol} can be interpreted by Fig.~\ref{fig:solution}. As shown in Appendix~\ref{pf:min_mse}, the condition of $\frac{\nu}{4^b} = \alpha \beta t_b^2 \exp(\beta t_b)$ should be satisfied for any $t_b > \delta$ (i.e., $\frac{\nu}{4^b} > \alpha \beta \delta^2 \exp(\beta \delta)$). If $\frac{\nu}{4^b} < \alpha \beta \delta^2 \exp(\beta \delta)$, then the corresponding refresh interval is forced to $t_b = \delta$. As the refresh power budget $\mathcal{P}$ decreases, the dual variable $\nu$ is increased to allocate longer refresh intervals. If more refresh power is available, then $\nu$ is lower and the corresponding refresh intervals are reduced as shown in Fig.~\ref{fig:solution}. 
	
	Note that $\vect{t}_0 = (\delta, \ldots, \delta)$ corresponds to the maximum refresh power and the minimum MSE as follows. 
	
	\begin{remark}[Maximum Refresh Power]\label{rem:max_power} The maximum refresh power is given by
		\begin{equation}
		\mathsf{P_{max}}=\mathsf{P}=(\vect{t}_0) = \frac{B}{\delta}. 
		\end{equation}
		If $B=8$ and $\delta = 0.064$, then $\mathsf{P_{max}} =125$.  	
	\end{remark}
	
	\begin{remark}[Minimum MSE]\label{rem:min_mse} The minimum MSE is 
		\begin{equation}
		\mathsf{MSE_{min}} = \mathsf{MSE}(\vect{t}_0) = \frac{4^B-1}{3} \cdot \alpha \exp(\beta \delta)
		\end{equation}
		which is obtained by the maximum refresh power. Note that the MSE increases exponentially with the refresh interval $\delta$. 
	\end{remark} 
	
	\subsection{Discrete Refresh Intervals}\label{sec:discrete}
	
	In the previous subsection, we formulated the convex optimization problem by assuming that any real values can be assigned to refresh intervals. Here, we investigate the discrete-valued refresh interval optimization. If the optimized discrete refresh intervals are multiples of $\delta$ (e.g., \SI{64}{\milli\second}), then the proposed optimization technique is compatible with current DRAM products. The reason is that any multiple of $\delta$ can be set as a refresh interval by gating the refresh commands~\cite{Ohsawa1998optimizing,Liu2012raidr}.  
	
	Suppose that $t_b = \Delta \cdot z_b$ where $\Delta = \gamma \delta$ and $z_b \in \mathbb{N}$ ($\mathbb{N}$ denotes the positive integers) for $b \in [0, B-1]$. Note that the step size of the refresh interval $\Delta$ is determined by $\gamma \in \mathbb{N}$, which controls the discrete optimization complexity and accuracy. Then, the convex optimization problem \eqref{eq:min_mse} can be modified into the following convex integer programming problem: 
	\begin{equation}
	\begin{aligned} \label{eq:min_mse_discrete}
	& \underset{\vect{z}}{\text{minimize}}
	& & \mathsf{MSE}(\vect{z}) = \sum_{b=0}^{B-1}{4^b \alpha \exp(\beta \gamma \delta \cdot z_b)}  \\
	&{\text{subject~to}} & & \mathsf{P}(\vect{z}) = \frac{1}{\gamma \delta} \sum_{b=0}^{B-1}{\frac{1}{z_b}} \le \mathcal{P} \\
	& & & z_b \in \mathbb{N}, \quad b=0,\ldots,B-1
	\end{aligned}
	\end{equation}
	where the positive integer solution $\vect{z}^{*}$ results in the optimized discrete refresh interval by $\widetilde{\vect{t}}^{*} = \Delta \cdot \vect{z}^{*}$.   
	
	Although convex integer programming is NP-hard, it can be solved much more efficiently than general integer non-linear programming problems~\cite{Bonami2008algorithmic,Bonami2012algorithms}. We obtained the optimized discrete solutions by standard mixed-integer non-linear program (MINLP) solvers. The numerical results are provided in Section~\ref{sec:numerical}.  
	
	\section{Numerical Results}\label{sec:numerical}
	
	\begin{figure}[t]
		\centering
		\includegraphics[width=0.45\textwidth]{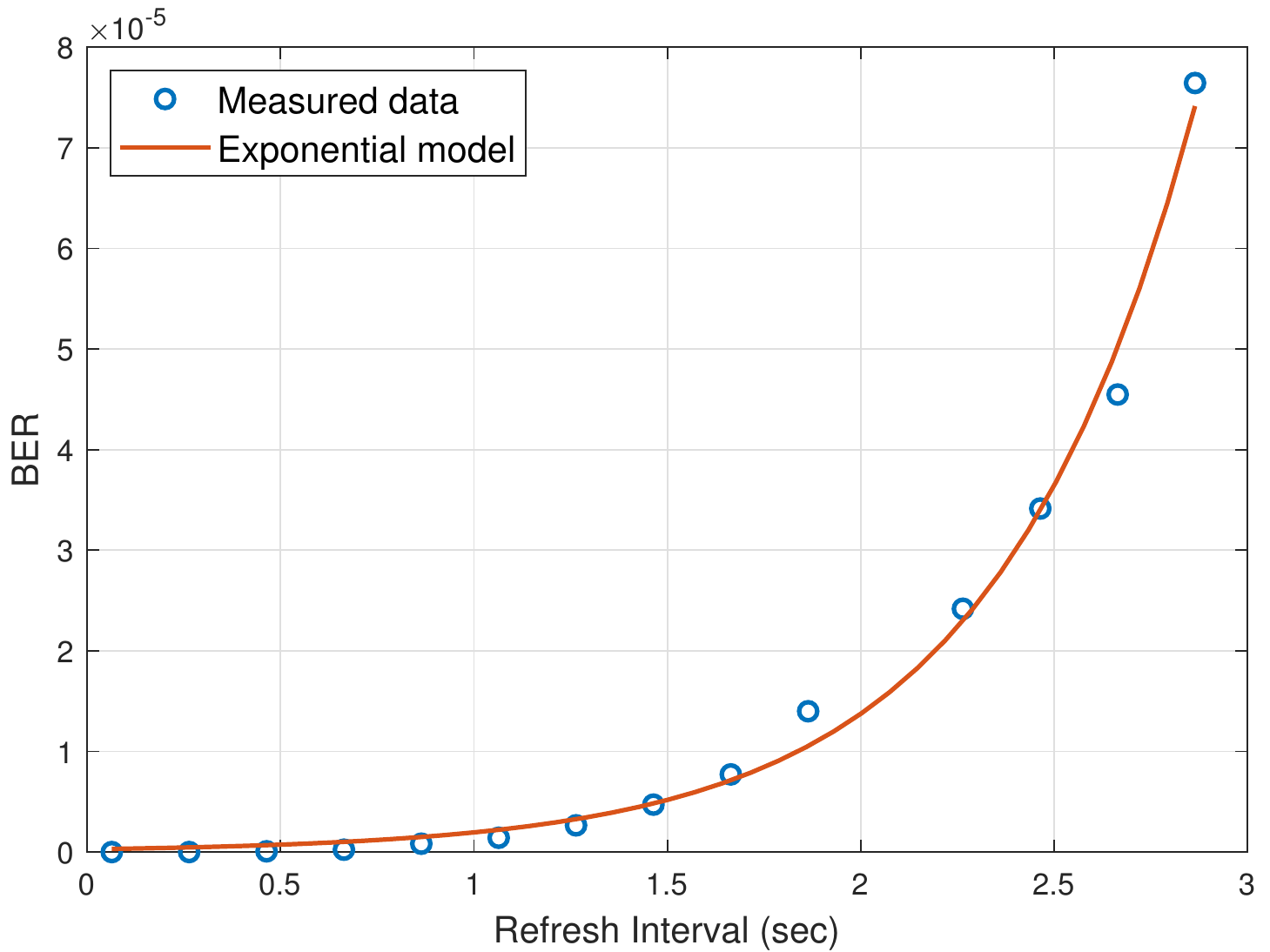}
		\caption{The measured BERs at \SI{80}{\celsius}~\cite[Table I]{Hou2013fpga} and the exponential model with the estimated $\alpha = 2.7737\times10^{-7}$ and $\beta=1.9508$.}
		\label{fig:ber_exponential}
		\vspace{-4mm}
	\end{figure}	
	
	\begin{figure}[t]
		\centering
		\subfloat[]{\includegraphics[width=0.45\textwidth]{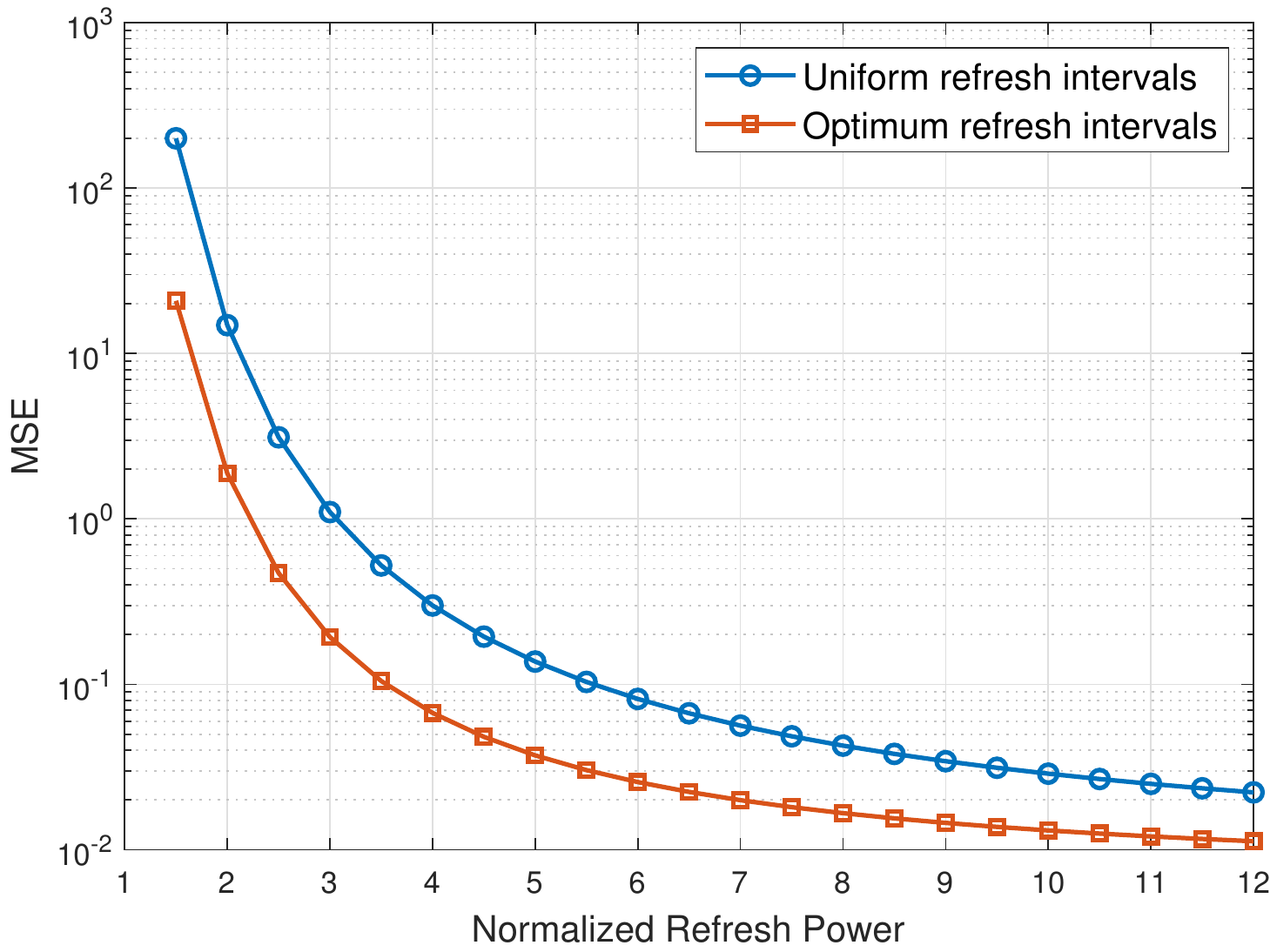}
			\label{fig:optimal_mse}}
		\hfil
		\subfloat[]{\includegraphics[width=0.45\textwidth]{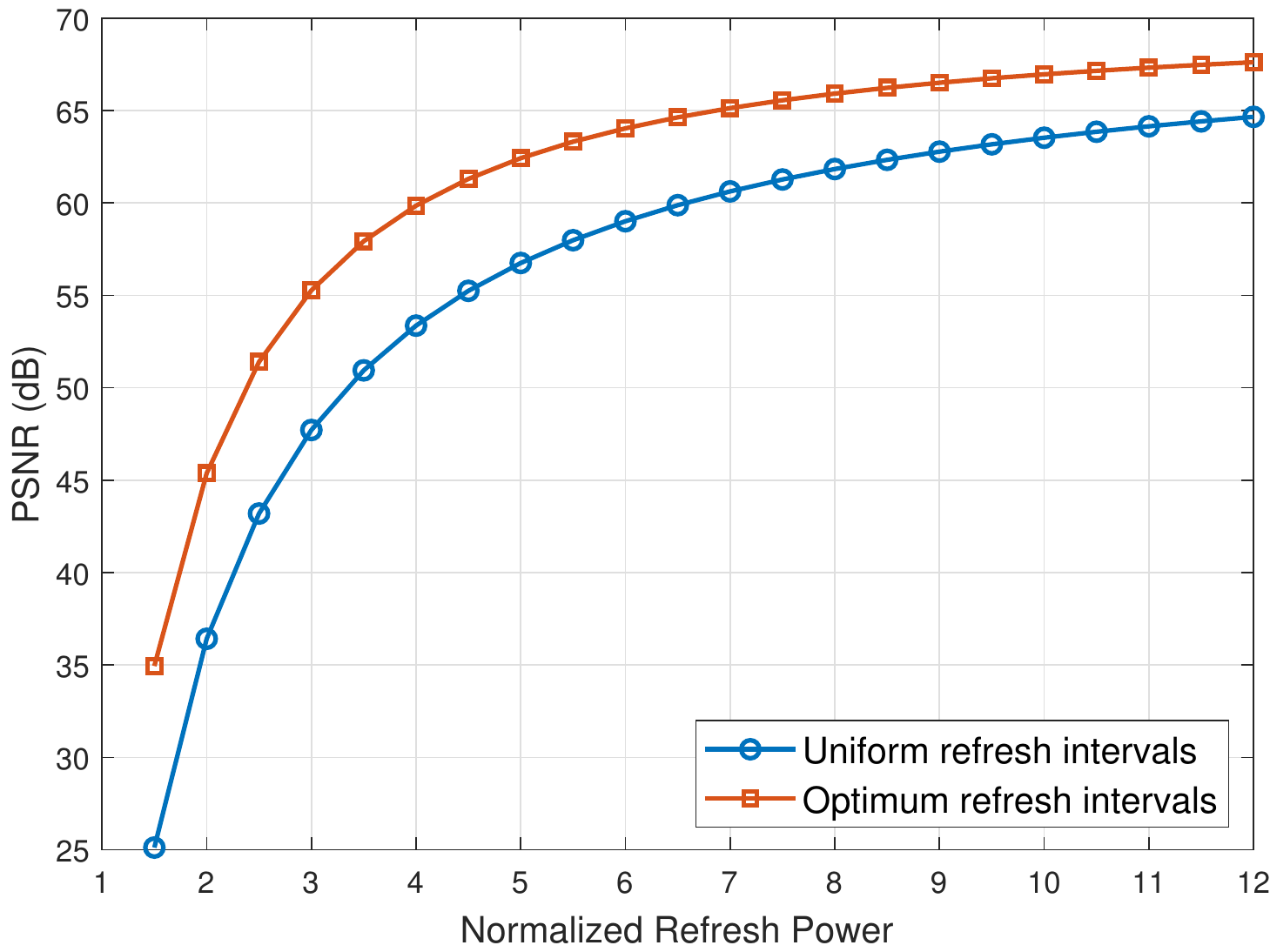}
			\vspace{-4mm}
			\label{fig:optimal_psnr}}
		\caption{Evaluation of proposed convex optimization in \eqref{eq:min_mse} (for $B = 8$): (a) MSE and (b) PNSR.}
		\label{fig:optimal}
		\vspace{-4mm}
	\end{figure}

	We evaluate the solutions of convex optimization problem \eqref{eq:min_mse} and the discrete optimization problem \eqref{eq:min_mse_discrete}. First, we estimate the parameters $\alpha$ and $\beta$ of \eqref{eq:exponential}. From the measured data in~\cite{Hou2013fpga}, we obtained the estimates of $\alpha = 2.7737\times10^{-7}$ and $\beta=1.9508$ (see Fig.~\ref{fig:ber_exponential}). Note that these parameters depend on manufacturers, products, and temperature as shown in \cite[Fig.~4]{Khan2014efficacy}. We note that higher-capacity, later-generation DRAM devices suffer from more retention failures~\cite{Liu2013experimental,Khan2014efficacy}. 
	
	Fig.~\ref{fig:optimal} shows numerical results by solving \eqref{eq:min_mse}. Fig.~\ref{fig:optimal}\subref{fig:optimal_mse} compares the MSEs of uniform refresh intervals and the optimal refresh intervals. At $\mathsf{MSE} = 1$, the optimal refresh intervals reduce the refresh power consumption by \SI{27}{\%}. For lower MSE, we can save more refresh power (e.g., \SI{36}{\%} refresh power reduction at $\mathsf{MSE} = 10^{-1}$).
	
	Fig.~\ref{fig:optimal}\subref{fig:optimal_psnr} compares the peak signal-to-noise ratios (PSNRs) of refresh interval assignments, which is a widely used fidelity metric for image and video quality. The PSNR depends on the MSE as 
	\begin{equation}
	\mathsf{PSNR} = 10 \log_{10}{\frac{(2^B-1)^2}{\mathsf{MSE}}}.
	\end{equation} 
	At $\mathsf{PSNR} = \SI{50}{dB}$, the optimized refresh intervals can reduce the refresh power by \SI{29}{\%}. Further, the optimized refresh intervals achieve \SI{38}{\%} power reduction at $\mathsf{PSNR} = \SI{60}{dB}$. The improvement by the optimized refresh intervals increases for a higher fidelity requirement. If we achieve a target fidelity (e.g., $\mathsf{PSNR} = \SI{50}{dB}$ is a quite reliable value in real-world images~\cite{Amer2005fast}), we do not need to waste power by refreshing every \SI{64}{\milli\second}, which requires $\mathsf{P_{max}} = 125$ (see Remark~\ref{rem:max_power}). Note that the optimized refresh interval assignment achieves $\mathsf{PSNR}=\SI{50}{dB}$ with $\mathsf{P}(\vect{t}^*) = 2.4$, which is less than \SI{2}{\%} of $\mathsf{P_{max}}$.

	\begin{figure}[t]
		\centering
		\includegraphics[width=0.45\textwidth]{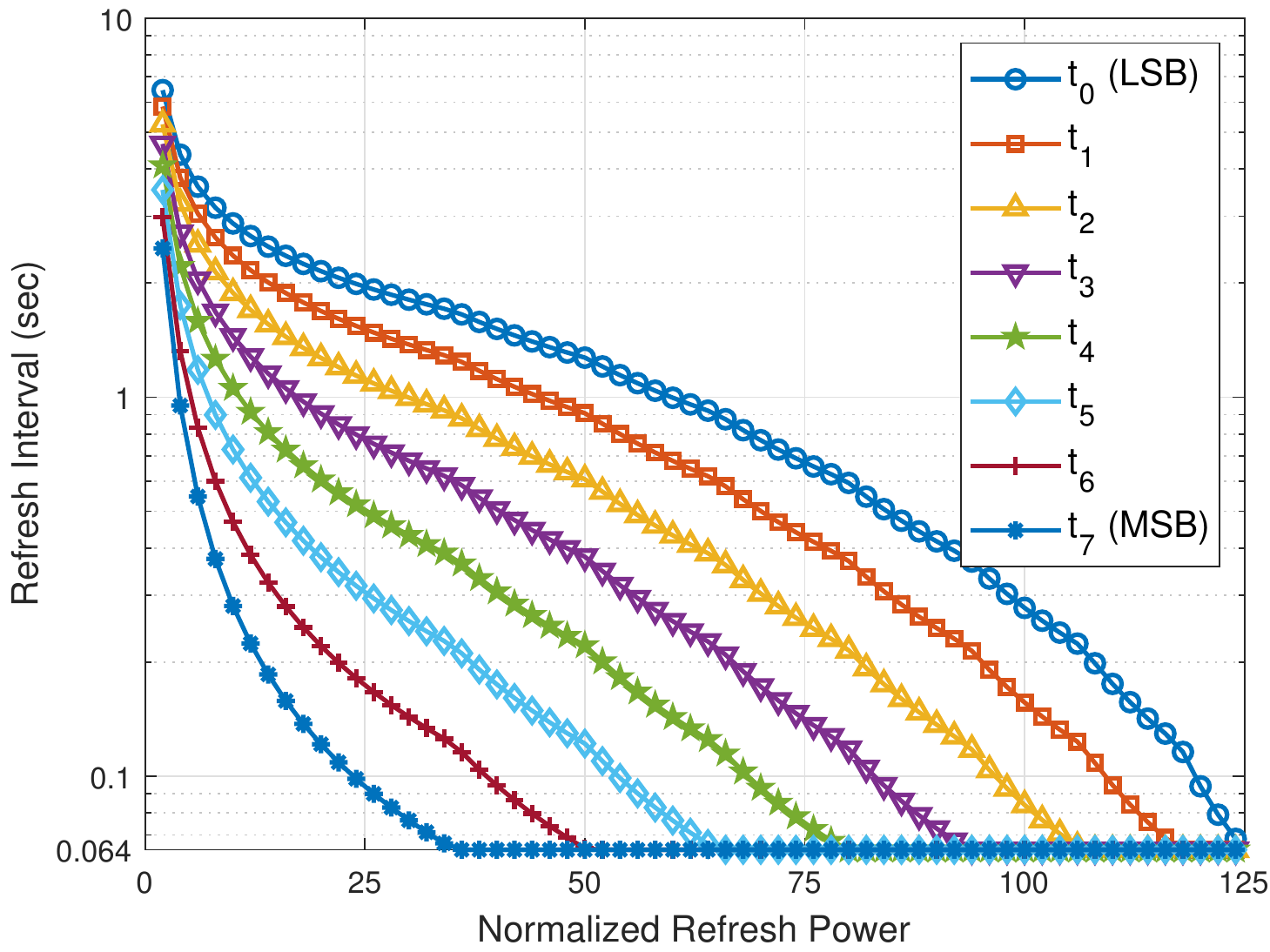}
		\caption{The optimal refresh interval assignments by Theorem~\ref{thm:min_mse}.}
		\label{fig:optimal_sol}
	\end{figure}

	Fig.~\ref{fig:optimal_sol} shows the optimal refresh interval assignments by Theorem~\ref{thm:min_mse}. The shorter refresh intervals (i.e., more refresh power assignments) are allocated to the more significant bits to minimize the MSE. As the refresh power budget $\mathcal{P}$ in \eqref{eq:min_mse} increases, the refresh intervals for more significant bits converge to $\delta$. Fig.~\ref{fig:optimal_sol} shows that $t_7 = \delta$ from $\mathcal{P} = 36$. More refresh intervals become $\delta$ for higher refresh power budget. 
	
\begin{figure}[t]
		\centering
		\includegraphics[width=0.45\textwidth]{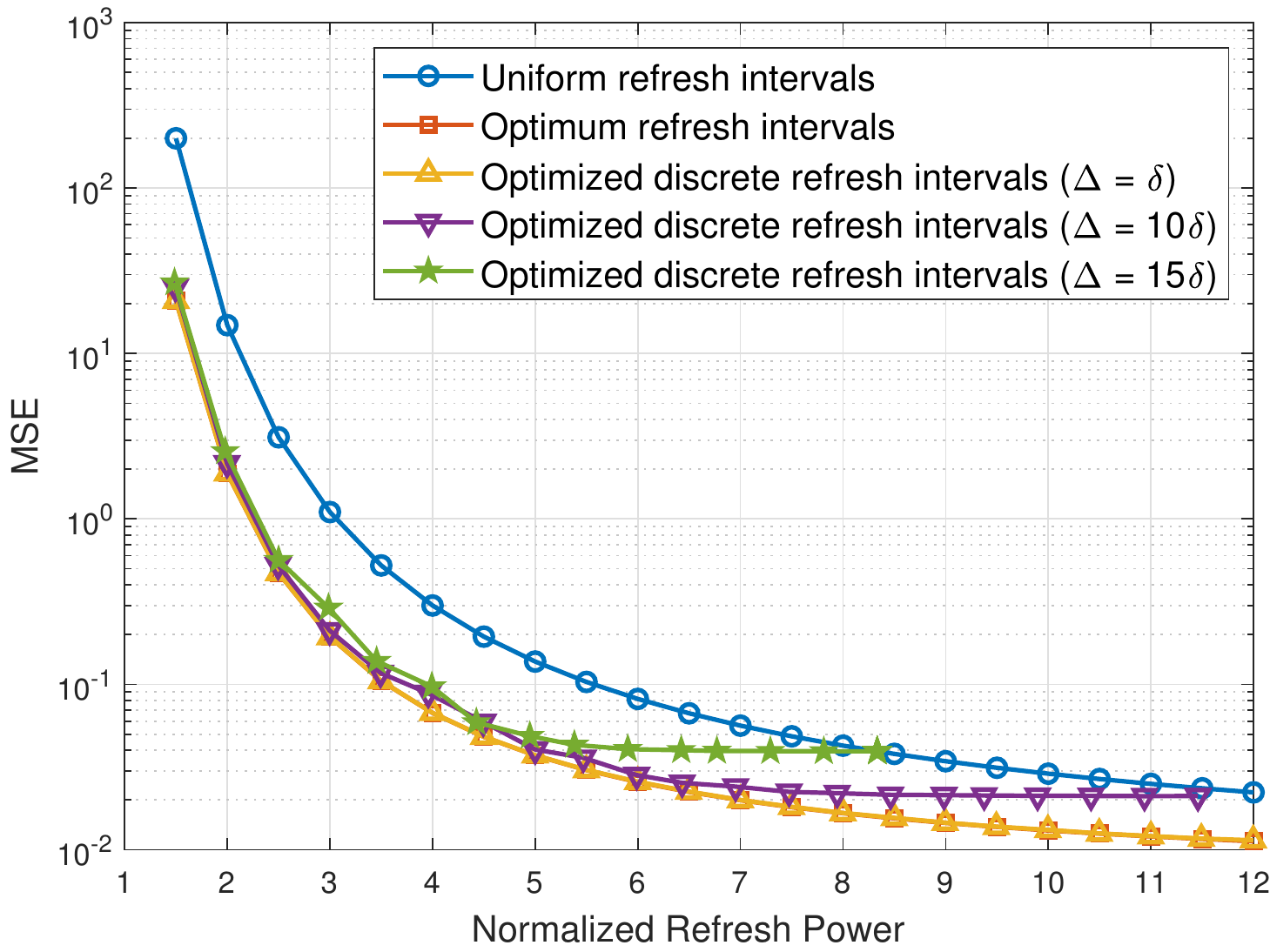}
		\caption{Evaluation of proposed discrete optimization of \eqref{eq:min_mse_discrete}.}
		\label{fig:discrete}
		\vspace{-4mm}
	\end{figure}

	Fig.~\ref{fig:discrete} shows the MSEs obtained by solving convex integer programming problem~\eqref{eq:min_mse_discrete}. This convex integer problem was solved by using \emph{Bonmin}~\cite{Bonami2012algorithms}. We observe that the MSE penalty due to discrete refresh intervals is negligible for a moderate step size $\Delta = \gamma \delta$. The MSE by discrete refresh intervals with $\Delta = \delta$ is almost the same as the optimal MSE. For $\Delta = 15 \delta$, the MSEs are distinct from the optimal MSEs from $\mathsf{P} = 6$. Note that the maximum refresh power with $\Delta = 15 \delta$ is $\mathsf{P} = \frac{B}{15 \delta} \simeq 8.33$.  
	
	\section{Conclusion}\label{sec:conclusion}
	
	We developed a principled approach to optimizing refresh intervals for energy-efficient memories. By formulating the convex optimization problem, we obtained the optimal refresh intervals to minimize the MSE under a refresh power budget. Also, we formulated a discrete optimization problem by taking into account the current DRAM standards and hardware implementation. The numerical results show that the optimum refresh intervals can achieve refresh power reductions of \SI{29}{\%} (at $\mathsf{PSNR} = \SI{50}{dB}$) and \SI{38}{\%} (at $\mathsf{PSNR} = \SI{60}{dB}$), respectively.


	\appendices
	
	\section{Proof of Theorem~\ref{thm:min_mse}}\label{pf:min_mse}
	
	The KKT conditions of \eqref{eq:min_mse} are as follows:
	\begin{align}
	\sum_{b=0}^{B-1}{\frac{1}{t_b}} &\le \mathcal{P}, \quad \nu \ge 0, \quad
	\nu \cdot \left(\sum_{b=0}^{B-1}{
		\frac{1}{t_b}} - \mathcal{P}\right) = 0, \label{eq:cr1_KKT_1} \\
	t_b &\ge \delta, \quad \lambda_b \ge 0, \quad \lambda_b \left( t_b - \delta \right) = 0 \label{eq:cr1_KKT_2} \\
	\frac{\partial L_1}{\partial t_b} &= 4^b \alpha \beta \exp(\beta t_b) - \frac{\nu}{t_b^2} - \lambda_b = 0 \label{eq:cr1_KKT_3}
	\end{align}
	for $b \in [0, B-1]$. From \eqref{eq:cr1_KKT_3}, $\lambda_b$ is given by
	\begin{equation} \label{eq:cr1_KKT_lambda}
	\lambda_b = 4^b \alpha \beta \exp(\beta t_b) - \frac{\nu}{t_b^2}. 
	\end{equation}
	From \eqref{eq:cr1_KKT_2} and \eqref{eq:cr1_KKT_lambda}, 
	\begin{equation}\label{eq:cr1_KKT_slack_1}
	\lambda_b \left( t_b - \delta \right) = \left(4^b \alpha \beta \exp(\beta t_b) - \frac{\nu}{t_b^2} \right) \left( t_b - \delta \right) = 0. 
	\end{equation}
	
	Suppose that $\nu = 0$. Then $\lambda_b = 4^b \alpha \beta \exp(\beta t_b) \ne 0$. Hence, $t_b = \delta$ for any $b \in [0, B-1]$. This is a trivial solution and the corresponding refresh power is $\mathsf{P}((\delta, \ldots, \delta))=\frac{B}{\delta}$. If this trivial solution does not violate the power budget constraint (i.e., $\frac{B}{\delta} \le \mathcal{P}$), then it will achieve the minimum MSE. However, we are more interested in the case of $\frac{B}{\delta} > \mathcal{P}$. Hence, we focus on $\nu \ne 0$, which results in $\sum_{b=0}^{B-1}{\frac{1}{t_b}} = \mathcal{P}$.   
	
	If $\lambda_b > 0$, then $t_b = \delta$. By~\eqref{eq:cr1_KKT_3}, the condition of $\lambda_b > 0$ is equivalent to $\frac{\nu}{4^b} < \alpha \beta t_b^2 \exp{(\beta t_b)}$. By~\eqref{eq:cr1_KKT_slack_1}, we claim that $t_b^* = \delta$ for $\frac{\nu}{4^b} < \alpha \beta \delta^2 \exp{(\beta \delta)}$. If $\lambda_b = 0$, then  
	\begin{equation}
	\alpha \beta t_b^2 \exp(\beta t_b) = \frac{\nu}{4^b}
	\end{equation} 
	which is equivalent to $\frac{\beta t_b}{2} \exp{\left(\frac{\beta t_b}{2}\right)} = \frac{\beta}{2}\sqrt{\frac{\nu}{4^b \alpha \beta}}$. By setting $x = \frac{\beta t_b}{2}$, we obtain $x \exp{(x)} = \frac{\beta}{2}\sqrt{\frac{\nu}{4^b \alpha \beta}}$. Hence, $W\left( \frac{\beta}{2}\sqrt{\frac{\nu}{4^b \alpha \beta}}\right) = x = \frac{\beta t_b}{2}$, i.e., $t_b = \frac{2}{\beta} W\left( \frac{\beta}{2} \sqrt{\frac{\nu}{4^b \alpha \beta}} \right)$.  
	
	
	%
	
	


\begin{thebibliography}{10}
	\bibitem{Bhati2016dram}
	I.~Bhati, M.~Chang, Z.~Chishti, S.~Lu, and B.~Jacob, ``{DRAM refresh
		mechanisms, penalties, and trade-offs},'' \emph{{IEEE} Trans. Comput.},
	vol.~65, no.~1, pp. 108--121, Jan. 2016.
	
	\bibitem{Emma2008rethinking}
	P.~G. Emma, W.~R. Reohr, and M.~Meterelliyoz, ``{Rethinking refresh: Increasing
		availability and reducing power in DRAM for cache applications},''
	\emph{{IEEE} Micro}, vol.~28, no.~6, pp. 47--56, Nov. 2008.
	
	\bibitem{Ohsawa1998optimizing}
	T.~Ohsawa, K.~Kai, and K.~Murakami, ``{Optimizing the DRAM refresh count for
		merged DRAM/logic LSIs},'' in \emph{Proc. {ACM/IEEE} Int. Symp. Low Power
		Electron. Design (ISLPED)}, Aug. 1998, pp. 82--87.
	
	\bibitem{Liu2012raidr}
	J.~Liu, B.~Jaiyen, R.~Veras, and O.~Mutlu, ``{RAIDR: Retention-aware
		intelligent DRAM refresh},'' in \emph{Proc. {ACM/IEEE} Annu. Int. Symp.
		Comput. Archit. (ISCA)}, Jun. 2012, pp. 1--12.
	
	\bibitem{Khan2014efficacy}
	S.~Khan, D.~Lee, Y.~Kim, A.~R. Alameldeen, C.~Wilkerson, and O.~Mutlu, ``The
	efficacy of error mitigation techniques for dram retention failures: A
	comparative experimental study,'' \emph{SIGMETRICS Perform. Eval. Rev.},
	vol.~42, no.~1, pp. 519--532, Jun. 2014.
	
	\bibitem{Liu2011flikker}
	S.~Liu, K.~Pattabiraman, T.~Moscibroda, and B.~G. Zorn, ``{Flikker: Saving DRAM
		refresh-power through critical data partitioning},'' \emph{SIGARCH Comput.
		Archit. News}, vol.~39, no.~1, pp. 213--224, Mar. 2011.
	
	\bibitem{Ghosh2007smart}
	M.~Ghosh and H.-H.~S. Lee, ``{Smart refresh: An enhanced memory controller
		design for reducing energy in conventional and 3D die-stacked DRAMs},'' in
	\emph{Proc. {IEEE/ACM} Annu. Int. Symp. Microarchitecture (MICRO)}, Dec.
	2007, pp. 134--145.
	
	\bibitem{Katayama1999fault}
	Y.~Katayama, E.~J. Stuckey, S.~Morioka, and Z.~Wu, ``{Fault-tolerant refresh
		power reduction of DRAMs for quasi-nonvolatile data retention},'' in
	\emph{Proc. IEEE Int. Symp. Defect and Fault Tolerance in VLSI Syst.}, Nov.
	1999, pp. 311--318.
	
	\bibitem{Wilkerson2010reducing}
	C.~Wilkerson, A.~R. Alameldeen, Z.~Chishti, W.~Wu, D.~Somasekhar, and S.-l. Lu,
	``{Reducing cache power with low-cost, multi-bit error-correcting codes},''
	in \emph{Proc. {ACM/IEEE} Annu. Int. Symp. Comput. Archit. (ISCA)}, Jun.
	2010, pp. 83--93.
	
	\bibitem{Chou2015reducing}
	C.~Chou, P.~Nair, and M.~K. Qureshi, ``{Reducing refresh power in mobile
		devices with morphable ECC},'' in \emph{Proc. IEEE/IFIP Int. Conf. Dependable
		Syst. Netw.}, Jun. 2015, pp. 355--366.
	
	\bibitem{Cho2014edram}
	K.~Cho, Y.~Lee, Y.~H. Oh, G.-c. Hwang, and J.~W. Lee, ``{eDRAM-based
		tiered-reliability memory with applications to low-power frame buffers},'' in
	\emph{Proc. {ACM/IEEE} Int. Symp. Low Power Electron. Design (ISLPED)}, Aug.
	2014, pp. 333--338.
	
	\bibitem{Choi2018comprehensive}
	W.~H. Choi, M.~Lueker-Boden, M.~Grobis, N.~Robertson, and Z.~Bandic, ``{A
		comprehensive study on DDR4 MRAM and ReRAM power estimation using a
		parameterized NVM power calculator},'' in \emph{Proc. {IEEE} Int. Memory
		Workshop (IMW)}, May 2018, pp. 1--4.
	
	\bibitem{Smullen2011relaxing}
	C.~W. Smullen, V.~Mohan, A.~Nigam, S.~Gurumurthi, and M.~R. Stan, ``{Relaxing
		non-volatility for fast and energy-efficient STT-RAM caches},'' in
	\emph{Proc. {IEEE} Int. Symp. High Performance Comput. Architecture (HPCA)},
	Feb. 2011, pp. 50--61.
	
	\bibitem{Jog2012cache}
	A.~Jog, A.~K. Mishra, C.~Xu, Y.~Xie, V.~Narayanan, R.~Iyer, and C.~R. Das,
	``{Cache revive: Architecting volatile STT-RAM caches for enhanced
		performance in CMPs},'' in \emph{Proc. Des. Autom. Conf (DAC)}, Jun. 2012,
	pp. 243--252.
	
	\bibitem{Kim2018generalized}
	Y.~Kim, M.~Kang, L.~R. Varshney, and N.~R. Shanbhag, ``{Generalized
		water-filling for source-aware energy-efficient SRAMs},'' \emph{{IEEE} Trans.
		Commun.}, vol.~66, no.~10, pp. 4826--4841, Oct. 2018.
	
	\bibitem{Chang2016understanding}
	K.~K. Chang, A.~Kashyap, H.~Hassan, S.~Ghose, K.~Hsieh, D.~Lee, T.~Li,
	G.~Pekhimenko, S.~Khan, and O.~Mutlu, ``{Understanding latency variation in
		modern DRAM chips: Experimental characterization, analysis, and
		optimization},'' in \emph{Proc. {ACM} SIGMETRICS Int. Conf. Meas. Model.
		Comput. Syst.}, Jun. 2016, pp. 323--336.
	
	\bibitem{Liu2013experimental}
	J.~Liu, B.~Jaiyen, Y.~Kim, C.~Wilkerson, and O.~Mutlu, ``{An experimental study
		of data retention behavior in modern DRAM devices: implications for retention
		time profiling mechanisms},'' in \emph{Proc. {ACM/IEEE} Annu. Int. Symp.
		Comput. Archit. (ISCA)}, Jun. 2013, pp. 60--71.
	
	\bibitem{Kim2003block}
	J.~Kim and M.~C. Papaefthymiou, ``{Block-based multiperiod dynamic memory
		design for low data-retention power},'' \emph{{IEEE} Trans. {VLSI} Syst.},
	vol.~11, no.~6, pp. 1006--1018, Dec. 2003.
	
	\bibitem{Kim2018sram}
	Y.~Kim, M.~Kang, L.~R. Varshney, and N.~R. Shanbhag, ``{SRAM bit-line swings
		optimization using generalized waterfilling},'' in \emph{Proc. IEEE Int.
		Symp. Inf. Theory (ISIT)}, Jun. 2018, pp. 1670--1674.
	
	\bibitem{Yang2011unequal}
	X.~Yang and K.~Mohanram, ``{Unequal-error-protection codes in SRAMs for mobile
		multimedia applications},'' in \emph{Proc. {IEEE/ACM} Int. Conf.
		Comput.-Aided Design (ICCAD)}, Nov. 2011, pp. 21--27.
	
	\bibitem{Hou2013fpga}
	C.~Hou, J.~Li, C.~Lo, D.~Kwai, Y.~Chou, and C.~Wu, ``{An FPGA-based test
		platform for analyzing data retention time distribution of DRAMs},'' in
	\emph{Proc. Int. Symp. VLSI Design, Autom., and Test}, Apr. 2013, pp. 1--4.
	
	\bibitem{Palomar2005practical}
	D.~P. Palomar and J.~R. Fonollosa, ``{Practical algorithms for a family of
		waterfilling solutions},'' \emph{{IEEE} Trans. Signal Process.}, vol.~53,
	no.~2, pp. 686--695, Feb. 2005.
	
	\bibitem{Corless1996lambert}
	R.~M. Corless, G.~H. Gonnet, D.~E.~G. Hare, D.~J. Jeffrey, and D.~E. Knuth,
	``{On the Lambert W function},'' \emph{Adv. Comput. Math.}, vol.~5, no.~1,
	pp. 329--359, Dec. 1996.
	
	\bibitem{Bonami2008algorithmic}
	P.~Bonami, L.~T. Biegler, A.~R. Conn, G.~Cornu{\'{e}}jols, I.~E. Grossmann,
	C.~D. Laird, J.~Lee, A.~Lodi, F.~Margot, N.~Sawaya, and A.~W{\"{a}}chter,
	``{An algorithmic framework for convex mixed integer nonlinear programs},''
	\emph{Discrete Optimization}, vol.~5, no.~2, pp. 186--204, May 2008.
	
	\bibitem{Bonami2012algorithms}
	P.~Bonami, M.~Kilin{\c{c}}, and J.~Linderoth, ``{Algorithms and software for
		convex mixed integer nonlinear programs},'' in \emph{Proc. Mixed Integer
		Nonlinear Programming}, Nov. 2012, pp. 1--39.
	
	\bibitem{Amer2005fast}
	A.~Amer and E.~Dubois, ``{Fast and reliable structure-oriented video noise
		estimation},'' \emph{{IEEE} Trans. Circuits Syst. Video Technol.}, vol.~15,
	no.~1, pp. 113--118, Jan. 2005.	
\end{thebibliography}
\end{document}